\documentstyle[preprint,aps]{revtex}

\begin{document}

\draft

\title{Mirror matter admixtures and isospin breaking in the $|\Delta I|=1/2$
rule in $\Omega^-$ two body non-leptonic decays}

\author{
G.~S\'anchez-Col\'on\cite{email1},
R.~Huerta
}
\address{
Departamento de F\'{\i}sica Aplicada.\\
Centro de Investigaci\'on y de Estudios Avanzados del IPN. Unidad
M\'erida.\\
A.P. 73, Cordemex. M\'erida, Yucat\'an, 97310. MEXICO.
}

\author{
A.~Garc\'{\i}a
}
\address{
Departamento de F\'{\i}sica.\\
Centro de Investigaci\'on y de Estudios Avanzados del IPN.\\
A.P. 14-740. M\'exico, D.F., 07000. MEXICO.
}

\date{\today}

\maketitle

\begin{abstract}
We discuss a description of $\Omega^-$ two body non-leptonic decays based
on possible, albeit tiny, admixtures of mirror matter in ordinary hadrons. The
$|\Delta I|=1/2$ rule enhancement is obtained as a result of isospin symmetry
and, more importantly, the rather large observed deviations from this rule
result from small isospin breaking. This analysis lends support to the
possibility that the enhancement phenomenon observed in low energy weak
interactions may be systematically described by mirror matter admixtures in
ordinary hadrons.
\end{abstract}


The enormous gap that exists from strong and electromagnetic interactions to
weak interactions offers a rather unique opportunity to either put very
stringent lower bounds on or to indirectly detect effects of the existence of
new matter, through its mixing with ordinary one. One attractive form of this
new matter may be mirror matter, initially discussed by Lee and Yang~\cite{lee}
in their seminal paper on parity violation and later systematically studied by
many authors~\cite{mohapatra}. Of particular interest is the possibility that
small admixtures of mirror hadrons in the ordinary ones may help describe the
enhancement phenomenon and its accompanying $|\Delta I|=1/2$ rule observed in
non-leptonic and weak radiative decays of the latter. After so many years of
its discovery~\cite{gellmann} this phenomenon still awaits a thorough
description. If mirror matter admixtures are to help in this respect, one
should demand that it does so in a systematic way and in terms of only a few
mixing angles that show a universality property. Otherwise it would be of
little help.

We have studied this possibility in a series of publications~\cite{wnlm}. To
implement such mixings it is necessary to introduce an ansatz, because no one is
yet able to perform the necessary QCD first principle calculations that enable
one to obtain such mixings starting at the quark level. In terms of that ansatz
one obtains results that comply with the above demands. A satisfactory
description for non-leptonic and weak radiative decays of hyperons and
non-leptonic decays of pseudoscalar mesons is obtained. The mixing angles come
out very small, albeit tiny, at around $10^{-6}$ - $10^{-7}$. It is no surprise
then that a very high lower bound on the mass of mirror hadrons can be set, at
around $10^6{\rm GeV}$~\cite{bound}. The origin of these numbers is the gap
mentioned in the beginning. Recently, under a separate cover~\cite{omega}, we
extended this approach to the two-body non-leptonic decays of $\Omega^-$. The
$|\Delta I|=1/2$ rule is obtained as a result of isospin symmetry and the
comparison with experiment is encouraging. However, a precise description of the
data was not yet obtained, because the isospin symmetry limit was assumed.

It is the purpose of this paper to study the effects of such breaking on the
mirror admixtures in $\Omega^-$ non-leptonic decays. This is delicate because
experimentally the $|\Delta I|=1/2$ rule in the modes $\Omega^- \to \Xi^-
\pi^0$ and $\Omega^-\to\Xi^0\pi^-$ is obeyed only to around $35\%$ in the
quotient of the corresponding branching ratios and in this approach this
deviation can be attributed only to isospin breaking. This should be contrasted
with $W$-mediated non-leptonic decays. Violations of this rule there are
attributed to a large $|\Delta I|=3/2$ piece in the effective hamiltonian. Here
we must explain a $35\%$ deviation with small isospin breaking. We shall see
this to be the case.

\paragraph*{Amplitudes of $\Omega^-$ decays with isospin breaking.}
As a start, let us briefly review the results of Ref.~\cite{omega}. Following
the ansatz discussed earlier~\cite{wnlm}, the mirror admixtures in $\Omega^-$
are

\begin{equation}
\Omega^-_{ph} =
\Omega^-_{s} -
\sqrt{3}\sigma\Xi^{*-}_{s} +
\sqrt{3}\delta'\Xi^{*-}_{p} +
\cdots.
\label{omega}
\end{equation}

\noindent
Our phase conventions are those of Ref.~\cite{gibson}.

Eq.~(\ref{omega}) is to be used together with the expressions with the mixings
in the $s=1/2$ baryons and pseudoscalar mesons relevant here, namely,

\begin{equation}
\Xi^-_{ph} =
\Xi^-_s -
\sigma\Sigma^-_s +
\delta'\Sigma^-_p
+ \cdots
\label{xi-}
\end{equation}

\begin{equation}
\Xi^0_{ph} =
\Xi^0_s -
\sigma
(\frac{1}{\sqrt 2}\Sigma^0_s + \sqrt{\frac{3}{2}}\Lambda_s) +
\delta'
(\frac{1}{\sqrt 2}\Sigma^0_p + \sqrt{\frac{3}{2}}\Lambda_p)
+ \cdots
\label{xi0}
\end{equation}

\begin{equation}
\Lambda_{ph} =
\Lambda_s +
\sigma\sqrt{\frac{3}{2}}(\Xi^0_s - n_s) +
\delta\sqrt{\frac{3}{2}}\Xi^0_p +
\delta'\sqrt{\frac{3}{2}}n_p
+ \cdots
\label{lambda}
\end{equation}

\begin{equation}
\pi^0_{ph} =
\pi^0_{p} -
\sigma\frac{1}{\sqrt 2}( K^0_{p} + \bar K^0_{p} ) +
\delta\frac{1}{\sqrt 2}( K^0_{s}- \bar K^0_{s} ) + \cdots
\label{pi0}
\end{equation}

\begin{equation}
\pi^-_{ph} =
\pi^-_{p} + \sigma K^-_{p} + \delta K^-_{s} + \cdots
\label{pi-}
\end{equation}

\begin{equation}
K^-_{ph} =
K^-_p -
\sigma \pi^-_p +
\delta' \pi^-_s +
\cdots
\label{k-}
\end{equation}

\noindent
The transition operator is the strong-interaction flavor and parity-conserving
hamiltonian responsible for the two-body strong decays of the other $s=3/2$
resonances in the decuplet where $\Omega^-_s$ belongs to. The parity-conserving
and parity-violating amplitudes ${\cal B}$ and ${\cal C}$ in the decay amplitude
$\bar{u}(p') ( {\cal B} + \gamma^5 {\cal C} ) q^{\mu} u_{\mu}(p)$ of $\Omega^-$
(in a standard notation) are given by

\begin{equation}
{\cal B}(\Omega^-\to\Xi^-\pi^0) =
-\sigma
(
 \sqrt 3 g_{{}_{\pi^0\Xi^-,\Xi^{*-}}} +
\frac{1}{\sqrt{2}}g_{{}_{\bar{K}^0\Xi^-,\Omega^-}}
) ,
\label{bxi-}
\end{equation}

\begin{equation}
{\cal C}(\Omega^-\to\Xi^-\pi^0) =
\delta'
\sqrt{3} g_{{}_{\pi^0_p\Xi^-_s,\Xi^{*-}_p}} -
\delta\frac{1}{\sqrt{2}}g_{{}_{\bar{K}^0_s\Xi^-_s,\Omega^-_s}}
,
\label{cxi-}
\end{equation}

\begin{equation}
{\cal B}(\Omega^-\to\Xi^0\pi^-) =
\sigma
(
- \sqrt 3 g_{{}_{\pi^-\Xi^0,\Xi^{*-}}} +
g_{{}_{K^-\Xi^0,\Omega^-}}
) ,
\label{bxi0}
\end{equation}

\begin{equation}
{\cal C}(\Omega^-\to\Xi^0\pi^-) =
\delta'
\sqrt{3} g_{{}_{\pi^-_p\Xi^0_s,\Xi^{*-}_p}} +
\delta g_{{}_{K^-_s\Xi^0_s,\Omega^-_s}}
,
\label{cxi0}
\end{equation}

\begin{equation}
{\cal B}(\Omega^-\to\Lambda K^-) =
\sigma
(
- \sqrt 3 g_{{}_{K^-\Lambda,\Xi^{*-}}} +
\sqrt{\frac{3}{2}}g_{{}_{K^-\Xi^0\Omega^-}}
) ,
\label{bl}
\end{equation}

\begin{equation}
{\cal C}(\Omega^-\to\Lambda K^-) =
\delta'
\sqrt{3} g_{{}_{K^-_p\Lambda_s,\Xi^{*-}_p}} +
\delta
\sqrt{\frac{3}{2}}g_{{}_{K^-_p\Xi^0_p,\Omega^-_s}}
.
\label{cl}
\end{equation}

\noindent
The $g_{{}_{MB',B}}$ in these amplitudes are the Yukawa coupling constants
observed in the strong decays $B \to B'M$. The constants
$g_{{}_{M_pB'_s,B_p}}$,  $g_{{}_{M_sB'_s,B_s}}$ and $g_{{}_{M_pB'_p,B_s}}$ are
new, because they involve mirror matter. The subindeces $s$ and $p$
stand for positive and negative parity, respectively. Our assumptions about the
$SU(3)$ properties of mirror $s=3/2$ resonances require that the
absolute values of their Yukawa couplings be the same as the corresponding ones
of ordinary $s=3/2$ resonances. However, their phases may differ.

We have no assumed in Eqs.~(\ref{bxi-}) -- (\ref{cl}) isospin symmetry. For
comparison purposes let us reproduce in Tables~\ref{table1} and \ref{table2}
the predictions obtained when one assumes this limit, when the $|\Delta I|=1/2$
rule is valid. One can see in these tables that $\Gamma(\Omega^- \to \Xi^-
\pi^0)$ and $\Gamma(\Omega^-\to\Xi^0\pi^-)$ differ by $16\%$ and $12\%$,
respectively, from their measured values. These two deviations give practically
all of the $\chi^2$ of 25.16. Their ratio or rather its inverse is 2.07 and is
smaller about $35\%$ than the measured one of 2.74. In this mirror admixture
approach these deviations must be explained by isospin breaking. One must recall
that in this approach the $|\Delta I|=1/2$ rule should more properly be named a
$\Delta I = 0$ rule.

\paragraph*{Effects of isospin breaking in the predictions for $\Omega^-$
decays.}
Since we do not assume isospin symmetry the constants
$g_{{}_{\pi^0\Xi^-,\Xi^{*-}}}$ and $g_{{}_{\pi^-\Xi^0,\Xi^{*-}}}$ and
$g_{{}_{\bar{K}^0\Xi^-,\Omega^-}}$ and $g_{{}_{K^-\Xi^0,\Omega^-}}$ are allowed
to vary separately, because they are no longer constrained to obey the $SU(2)$
symmetry relations $g_{{}_{\pi^-\Xi^0,\Xi^{*-}}} = - \sqrt{2}
g_{{}_{\pi^0\Xi^-,\Xi^{*-}}}$ and $g_{{}_{K^-\Xi^0,\Omega^-}} =
g_{{}_{\bar{K}^0\Xi^-,\Omega^-}}$. The results are also displayed in
Tables~\ref{table1} and \ref{table2}. The main changes appear in
$\Gamma(\Omega^- \to\Xi^- \pi^0)$ and $\Gamma(\Omega^-\to\Xi^0\pi^-)$ and their
corresponding parity-conserving amplitudes ${\cal B}(\Omega^- \to \Xi^- \pi^0)$
and ${\cal B}(\Omega^- \to \Xi^0 \pi^-)$. Percentage-wise these changes
are $-15.4\%$, $+11.4\%$, $-8.0\%$, and $+5.6\%$, respectively. All percent
changes we shall quote are with respect to the symmetry limit values. Changes in
other observables and amplitudes are not perceptible at the two and even at the
three digit level. With isospin breaking the description of the data is quite
satisfactory.

To appreciate this breaking we must look first at the values of the parameters.
Our $\chi^2$ consists of 10 restrictions, 3 weak decay rates, 3 asymmetries, 1
strong decay rate~\cite{pdg} and the values of the 3 angles from earlier
work~\cite{wnlm}. The latter are $\sigma = (4.9\pm 2.0) \times 10^{-6}$,
$\delta = (2.2\pm 0.9)\times 10^{-7}$, and $\delta' = (2.6\pm 0.9)\times
10^{-7}$.  In the $SU(2)$ limit, the 6 parameters take the values
$g_{{}_{\pi^0\Xi^-,\Xi^{*-}}} = 4.326{\rm GeV}^{-1}
\ (= g_{{}_{\pi^0_p\Xi^-_s,\Xi^{*-}_p}})$,
$g_{{}_{\pi^-\Xi^0,\Xi^{*-}}} = - 6.118{\rm GeV}^{-1}
\ (= g_{{}_{\pi^-_p\Xi^0_s,\Xi^{*-}_p}})$,
$g_{{}_{\bar{K}^0\Xi^-,\Omega^-}} = - 10.35{\rm GeV}^{-1}
\ (= - g_{{}_{\bar{K}^0_s\Xi^-_s,\Omega^-_s}})$,
$g_{{}_{K^-\Xi^0,\Omega^-}} = - 10.35{\rm GeV}^{-1}
\ (= - g_{{}_{K^-_s\Xi^0_s,\Omega^-_s}}
= g_{{}_{K^-_p\Xi^0_p,\Omega^-_s}})$,
$g_{{}_{K^-\Lambda,\Xi^{*-}}} = - 7.773{\rm GeV}^{-1}
\ (= - g_{{}_{K^-_p\Lambda_s,\Xi^{*-}_p}})$,
$\sigma = 5.10 \times 10^{-6}$, $\delta = 2.63 \times 10^{-7}$, and $\delta' =
2.15 \times 10^{-7}$. The second and fourth constants are fixed by $SU(2)$ and
we have displayed in parentheses the phases used for the coupling constants
involving mirror hadrons.

When isospin is broken the 8 parameters used take the values
$g_{{}_{\pi^0\Xi^-,\Xi^{*-}}} = 4.322{\rm GeV}^{-1}$,
$g_{{}_{\pi^-\Xi^0,\Xi^{*-}}} = - 6.120{\rm GeV}^{-1}$,
$g_{{}_{\bar{K}^0\Xi^-,\Omega^-}} = - 10.36{\rm GeV}^{-1}$,
$g_{{}_{K^-\Xi^0,\Omega^-}} = - 10.34{\rm GeV}^{-1}$,
$g_{{}_{K^-\Lambda,\Xi^{*-}}} = - 7.766{\rm GeV}^{-1}$,
$\sigma = 5.10 \times 10^{-6}$, $\delta = 2.62 \times 10^{-7}$, and $\delta' =
2.15 \times 10^{-7}$, with the same phases as above for the mirror couplings.
These coupling constants are of the order of magnitude expected~\cite{lopez}.
One can see that the angles remain unchanged from their symmetry limit to their
symmetry breaking predictions. The first four couplings in these two lists
change only in the third or even in the fourth digit. In percent, the changes in
the ratios $(- \sqrt{2})g_{{}_{\pi^0\Xi^-,\Xi^{*-}}}/g_{{}_{\pi^-\Xi^0,\Xi^{*-}}}$ and
$g_{{}_{K^-\Xi^0,\Omega^-}}/g_{{}_{\bar{K}^0\Xi^-,\Omega^-}}$ are $-0.13\%$, and
$-0.19\%$, respectively. This is indeed a small isospin breaking.

\paragraph*{Relevant remarks.}
It is interesting to see how such a small breaking leads to a deviation of
around $35\%$ in the quotient of the branching ratios of $\Omega^- \to \Xi^-
\pi^0$ and $\Omega^-\to\Xi^0\pi^-$. We must concentrate on the ${\cal B}$
amplitudes of these two decays, the ${\cal C}$ amplitudes are kinematically
suppressed and this explains the small values of the asymmetry coefficients.
These amplitudes are particularly sensitive to isospin $SU(2)$ symmetry and to
its breaking. An order of magnitude estimate shows this.

Looking at Eqs.~(\ref{bxi-}) -- (\ref{cl}) one sees that since the mixing angle
$\sigma$ is of the order of $10^{-5}$ - $10^{-6}$ and the Yukawa couplings are
of the order of 10 ${\rm GeV}^{-1}$, the amplitudes are expected to be of the
order of $10^{-7}$ - $10^{-8}$ ${\rm MeV}^{-1}$. In contrast, they are of the
order of $10^{-9}{\rm MeV}^{-1}$. From Table~\ref{table2}, ${\cal B}(\Omega^-
\to \Xi^- \pi^0) = - 0.8889\times 10^{-9}{\rm MeV}^{-1}$ and ${\cal B}(\Omega^-
\to \Xi^0 \pi^-) = 1.257\times 10^{-9}{\rm MeV}^{-1}$ in the symmetry limit. So,
experimentally our estimate must be reduced by over one order of magnitude.
This is what $SU(2)$ symmetry achieves, but then the ${\cal B}$ amplitudes
become very sensitive to its breaking. Accordingly, a small breaking of around
$0.20\%$ in the couplings lead to the new values ${\cal B}(\Omega^- \to \Xi^-
\pi^0) = - 0.8175\times 10^{-9}{\rm MeV}^{-1}$ and ${\cal B}(\Omega^- \to \Xi^0
\pi^-) = 1.327\times 10^{-9}{\rm MeV}^{-1}$. The corresponding percent changes
are $-8.0\%$ and $+5.6\%$. The ratios of these two amplitudes deviates around
$15\%$ from the $-1/\sqrt{2}$ value predicted by the $|\Delta I|=1/2$ rule. Since
the ${\cal B}$'s dominate the branching ratios, the quotient
$\Gamma(\Omega^-\to\Xi^0\pi^-)/\Gamma(\Omega^-\to\Xi^-\pi^0)$ deviates
around $35\%$ from its 2.1 value of the $|\Delta I|=1/2$ rule. A $-8.0\%$ or a
$+5.6\%$ change in the ${\cal B}$'s is large, and not because $SU(2)$ breaking
is large but because the ${\cal B}$'s are small.

$SU(2)$ symmetry becomes so relevant because of the demands we discussed above.
It is necessary that the values of the mixing angles used here be the same as in
earlier work and that the Yukawa couplings be of the right order of magnitude.
The role of the strong decay rate $\Gamma_4$ in Table~\ref{table1} is
essential. Without this restriction those couplings would become wild, and the
required suppression would be provided by a reduction of their magnitudes and
not by the interplay of the $SU(2)$ Clebsch-Gordan coefficients.

In Ref.~\cite{bound} we referred to the approach used here as manifest mirror
symmetry, because it is assumed that the strong and electromagnetic interactions
are shared with the same intensity between mirror and ordinary
matter~\cite{senjanovic}. These are the most favorable conditions under which
mirror admixtures may give observable effects in our by large predominantly
ordinary matter world.

\paragraph*{Discussion and conclusion.}
The above analysis together with the previous ones shows that mirror-matter
admixtures provide a systematic description of the enhancement phenomenon
observed in non-leptonic meson, hyperon, and $\Omega^-$ decays (NLMD, NLHD, and
NLOD) and in weak radiative decays of hyperons (WRD). At this point the
question is what room is left for those contributions by the description of
this phenomenon within the standard framework. We shall now address this
question.

Theoretical investigations of the enhancement phenomenon have a long story (see
Refs.~\cite{11,12,13,14,15,16,17,18,19,20,21} and references therein) and
progress has been slow. It is a common understanding that its explanation is
rooted in the dynamics of the strong interactions that dress the weak
transition vertex. At present there is not yet a first principle QCD derivation
of this effect. One must therefore resort to ways to address calculations, which
lead to effective weak hamiltonians involving quasi-particle degrees of freedom.
Necessarily, many parameters as low energy coupling constants or effective
masses are introduced, limiting the predictive power. In the last decade an a
half perturbative contributions have been shown to produce rather small
enhancement. However, they do provide essential hints of how non-perturbative
contributions should be incorporated in the low energy domain of QCD.
Currently, a consensus has been reached that the major effect in understanding
the $\Delta I=1/2$ rule and its deviations is due to non-perturbative long
distance contributions. However, comparison between different effective
approaches is difficult. There are ambiguities from author to author and
uncertainties in calculations may be largely amplified. In some cases (NLMD and
NLHD) an effort is made to estimate a theoretical error, in others (WRD and
NLOD) the predictions are less quantitative or simply qualitative.

Important attention has been paid to studying systematically the success of
specific approaches in different groups of decays, in order to assess
their overall success. The best numerical results in reproducing
experimental measurements are obtained in NLMD and NLHD. In Ref.~\cite{13}
NLMD are analyzed and after adjusting a free parameter corresponding to
unknown Fierz terms in factorization, the theoretical predictions carry a
20-30\% uncertainty and the central values differ from experiment by
25-30\%. Also in this reference~\cite{13} the results of \cite{14} for NLHD
are reviewed. Out of the 14 measured amplitudes 13 are reproduced between
1-20\%, only one is missed by about 80\%. The corresponding decay rates
are all reproduced ranging from 1\% to 30\% [14].The results of
Ref.~\cite{15} for NLHD reproduce 4 of the $P$-waves amplitudes (only 4 are
independent when exact isospin is assumed) very well between 1-5\%;
however, only 3 $S$-wave amplitudes are reproduced between 10-30\% and one
is missed badly (its sign is predicted opposite).

The approach of \cite{13,14} has not been applied to WRD. In Ref.~\cite{16}
WRD are studied, but its results are not quantitative. Experimental data
is used to extract some parameters, which are then compared with their
determination elsewhere. An order of magnitude agreement is observed. The
approach of \cite{15} has been extended to WRD \cite{17}; however, only
indicative results are expected. The large negative asymmetry of
$\Sigma^+\to p\gamma$ is reproduced, but the corresponding branching ratio
is missed by an order of magnitude. Predictions within the standard
framework for this group of decays are not yet quantitatively
satisfactory.

The approaches of \cite{13,14} and \cite{15} have both been applied to NLOD
\cite{14,18}; although the authors make it clear that their results are mainly
of qualitative nature and are not anticipated to reproduce precisely
experimental numbers. Looking at the predictions of \cite{14} one can see that
the three decay rates are predicted within 30-50\%. In \cite{18} numerical
results are better, at 5-10\%; however, the authors emphasized that these
results are not to be trusted yet. There are other predictions in the
literature \cite{19,20}, whose approach is along the lines of \cite{18}. Ref. \cite{19}
does not really make predictions, it is similar in spirit to \cite{16}, certain
parameters are fitted and their order of magnitude is satisfactorily
compared to independent determinations elsewhere. In contrast, Ref. \cite{20}
points out that the $\Delta I=3/2$ amplitudes in NLOD turn out so large as
to cast doubt on either experimental values or on the reliability of
approximations of calculations of the analogous amplitudes in NLHD. A
different effective approach is envisaged in Ref.~\cite{21}. Their
predictions for decay rates, however, turn out to be a factor of 2 off
experiment.
 
To summarize our brief review of the work within the standard framework,
let us say that (i) no first principle QCD calculations are yet available,
(ii) there are many effective formulations which are difficult to compare
with one another, however, (iii) predictions so far are encouraging and
(iv) a systematic understanding of long range effects to produce
enhancement is $\Delta S = 1$ non-leptonic hadronic decays is gradually
emerging. We can now address our question: is there room for contributions
of physics outside the standard framework in these decays? Currently the
answer is affirmative. Guiding ourselves by the estimated theoretical
errors and the accuracy of the predictions of Refs.~\cite{11,12,13,15,18}
one may estimate that the enhancement produced in the standard framework
leaves room for other contributions that ranges between 30-50\%.

In view of our results above and of previous work and of this last
discussion, one may conclude that in attacking the problem of the
enhancement observed in $\Delta S = 1$ non-leptonic and radiative decays
one should keep in mind that non-standard physics may contribute to this
enhancement. The extent of this contribution will depend on the room left
for it by standard framework contributions. Whether these latter will
produce 50\%, 70\%, or even a full 100\% of the observed enhancement will
be determined beyond doubt by first principle calculations of QCD long
range effects. This final answer is, however, not yet within our reach and
may stay so for quite some time \cite{11}. It is therefore very important
in the meanwhile to remain open-minded and not to overlook potentially
important contributions from outside the standard framework.

The authors are grateful to CONACyT (M\'exico) for partial support.

\begin{table}
\caption{Experimental decay rates and asymmetry coefficients and their
predicted values. The upper entries assume isospin symmetry, the lower ones
allow for breaking of the latter. The indices 1, 2, and 3 refer to the modes
$\Omega^- \to \Xi^-\pi^0$, $\Omega^- \to \Xi^0\pi^-$, and $\Omega^- \to \Lambda
K^-$, respectively. The index 4 refers to the total strong decay rates of
$\Xi^{*-}\to\Xi\pi = \Xi^{*-}\to\Xi^-\pi^0 + \Xi^{*-}\to\Xi^0\pi^-$}.
\label{table1}
\begin{tabular}{l c d d}
\hline
Decay & Experiment & Prediction & $\chi^2$ \\
\hline
$\Gamma_1(10^9{\rm seg}^{-1})$ & $1.046\pm 0.051$ & 1.241 & 14.64 \\
& & 1.050 & 0.006 \\
$\alpha_1$ & $0.05\pm 0.21$ & 0.076 & 0.02 \\
& & 0.082 & 0.02 \\
$\Gamma_2(10^9{\rm seg}^{-1})$ & $2.871\pm 0.095$ & 2.571 & 9.96 \\
& & 2.863 & 0.007 \\
$\alpha_2$ & $0.09\pm 0.14$ & 0.078 & 0.01 \\
& & 0.071 & 0.02 \\
$\Gamma_3(10^9{\rm seg}^{-1})$ & $8.25\pm 0.15$ & 8.247 & 0.0004 \\
& & 8.250 & 0.0 \\
$\alpha_3$ & $-$$0.026\pm 0.026$ & $-$0.021 & 0.04 \\
& & $-$0.020 & 0.06 \\
$\Gamma_4({\rm MeV})$ & $9.9\pm 1.9$ & 9.8 & 0.001 \\
& & 9.8 & 0.001 \\
\hline
\end{tabular}
\end{table}

\begin{table}
\caption{Predictions for the parity-conserving and parity-violating amplitudes
in $\Omega^-$ two-body non-leptonic decays assuming isospin symmetry limit and
not assuming it. The percent variatons are with respect to the symmetry limit
values. The indices are as in Table~I.}
\label{table2}
\begin{tabular}{c d d d}
\hline
Decay &
$SU(2)$ symmetry &
$SU(2)$ symmetry &
$\%$ variation \\
$(10^{-9}{\rm MeV}^{-1})$ &
limit &
breaking &
\\
\hline
${\cal B}_1$ &
$-$0.8889 &
$-$0.8175 &
$-$8.0 \\
${\cal C}_1$ &
$-$0.3138 &
$-$0.3098 &
$-$1.3 \\
${\cal B}_2$ &
1.257 &
1.327 &
$+$5.6 \\
${\cal C}_2$ &
0.4438 &
0.4300 &
$-$3.1 \\
${\cal B}_3$ &
4.014 &
4.015 &
$+$0.02 \\
${\cal C}_3$ &
$-$0.4392 &
$-$0.4259 &
$-$3.0 \\
\hline
\end{tabular}
\end{table}


\begin{references}

\bibitem[*]{email1}
e-mail address: gsanchez@mda.cinvestav.mx

\bibitem{lee}
T.~D.~Lee, and C.~N.~Yang,
Phys. Rev. {\bf 104} (1956) 254.

\bibitem{mohapatra}
R.~N.~Mohapatra,
{\it Unification and Supersymmetry. The frontiers of quark-lepton physics,}
Contemporary Physics
(Springer-Verlag, N.~Y., 1986).

\bibitem{gellmann}
M.~Gell-Mann and A.~Pais,
{\it Proc. Intern. Conf. High Energy Phys., Glasgow, 1954}
(Pergamon Press, London, 1955), p.342.

\bibitem{wnlm}
A.~Garc\'{\i}a, R.~Huerta, and G.~S\'anchez-Col\'on,
Rev.\ Mex.\ Fis.\ {\bf 43}\ (1997)\ 232;
J.\ Phys.\ G:\ Nucl.\ Part.\ Phys.\ {\bf 24}\ (1998)\ 1207;
{\bf 25}\ (1999)\ 45;
{\bf 25}\ (1999)\ L1;
{\bf 26}\ (2000)\ 1417;
Mod.\ Phys.\ Lett.\ {\bf A 15}\ (2000)\ 1749.

\bibitem{bound}
A.~Garc\'{\i}a, R.~Huerta, and G.~S\'anchez-Col\'on,
Phys.\ Lett.\ {\bf B 498}\ (2001)\ 251.

\bibitem{omega}
G.~S\'anchez-Col\'on, R.~Huerta, and A.~Garc\'{\i}a
(2002)
{\it Mirror matter admixtures in $\Omega^-$ two body non-leptonic decays and the
$|\Delta I|=1/2$ rule}
(hep-ph/0210316, submitted for publication).

\bibitem{gibson}
W.~M.~Gibson and B.~R.~Pollard,
{\it Symmetry Principles in Elementary Particle Physics}
(Cambridge: Cambridge University Press 1976).

\bibitem{pdg}
D.~E.~Groom {\it et al.}, Eur.\ Phys.\ J.\ {\bf C 15}\ (2000)\ 1.

\bibitem{lopez}
G.~Lopez Castro and A. Mariano, Nucl.\ Phys.\ {\bf A 697}\ (2002)\ 440.

\bibitem{senjanovic}
S.~M.~Barr, D.~Chang, and G.~Senjanovi\'c,
Phys.\ Rev.\ Lett.\ {\bf 67} (1991) 2765.

\bibitem{11}
S. Bertolini, hep-ph/0206095.

\bibitem{12}
E. Pallante and A. Pich,
Nucl.\ Phys.\ {\bf B 592}\ (2001)\ 294.

\bibitem{13}
M. Neubert and B. Stech,
Phys.\ Rev.\ {\bf D 44}\ (1991)\ 775.

\bibitem{14}
B. Stech and Q.P. Xu,
Z.\ Phys.\ {\bf C 49}\ (1991)\ 491.

\bibitem{15}
B. Borasoy and B.R. Holstein,
Phys.\ Rev.\ {\bf D 59}\ (1999)\ 094025.

\bibitem{16}
H. Neufeld,
Nucl.\ Phys.\ {\bf B 402}\ (1993)\ 166.

\bibitem{17}
B. Borasoy and B.H. Holstein,
Phys.\ Rev.\ {\bf D 59}\ (1999)\ 054019.

\bibitem{18}
B. Borasoy and B.H. Holstein,
Phys.\ Rev.\ {\bf D 60}\ (1999)\ 054021.

\bibitem{19}
D.A. Egolf, I.V. Melnikov, and R.P. Springer,
Phys.\ Lett.\ {\bf B 451}\ (1999)\ 267.

\bibitem{20}
J. Tandean and G. Valencia,
Phys.\ Lett.\ {\bf B 452}\ (1999)\ 395.

\bibitem{21}
G. Duplancic, H. Pasagic, M. Praszalowicz, and J. Trampetic,
Phys.\ Rev.\ {\bf D 65}\ (2002)\ 054001.

\end{references}
\end{document}